\documentclass[a4paper,11pt]{article}
\pdfoutput=1 
\usepackage{jcappub} 
\usepackage[T1]{fontenc}\usepackage[british]{babel}
\usepackage{xcolor}
\usepackage{booktabs}
\usepackage{latexsym}
\usepackage{amssymb}
\usepackage{amsmath}
\usepackage{graphicx}
\usepackage{tabularx}
\usepackage{cancel}
\usepackage{hyperref}
\usepackage{ulem}
\usepackage[]{cleveref}
\newcommand{\dd}{\text{d}}

\newcommand{\varpar}{{\mkern3mu\vphantom{\perp}\vrule depth 0pt\mkern2mu\vrule depth 0pt\mkern3mu}}

\title{\boldmath Quasi-local masses and cosmological coupling of black holes and  mimickers}

\author[a,b]{Mariano Cadoni,}
\author[c,d]{ Riccardo Murgia,}
\author[a,b]{ Mirko Pitzalis,}
\author[a,b,1]{ Andrea P. Sanna\note{Corresponding author.}}

\affiliation[a]{Dipartimento di Fisica, Universit\`a di Cagliari, Cittadella Universitaria, 09042 Monserrato, Italy}
\affiliation[b]{INFN, Sezione di Cagliari, Cittadella Universitaria, 09042 Monserrato, Italy}
\affiliation[c]{Gran Sasso Science Institute (GSSI), Viale F. Crispi 7, L'Aquila (AQ), I-67100, Italy}
\affiliation[d]{INFN - Laboratori Nazionali del Gran Sasso (LNGS), L'Aquila (AQ), I-67100, Italy}

\emailAdd{mariano.cadoni@ca.infn.it}
\emailAdd{riccardo.murgia@gssi.it}
\emailAdd{mirko.pitzalis@ca.infn.it}
\emailAdd{asanna@dsf.unica.it}

\abstract{Motivated by the recent heated debate on whether the masses of local objects, such as compact stars or black holes (BHs), may be affected by the large-scale, cosmological dynamics, we analyze the conditions under which, in a general relativity framework, such a coupling small/large scales is allowed. We shed light on some controversial arguments, which have been used to rule out the latter possibility. We find that the cosmological coupling occurs whenever the energy of the central objects is quantified by the quasi-local Misner-Sharp mass (MS). Conversely, the decoupling occurs whenever the MS mass is fully equivalent to the (nonlocal) Arnowitt-Deser-Misner (ADM) mass. Consequently, for singular BHs embedded in cosmological backgrounds, like the Schwarzschild-de Sitter or McVittie solutions,  we show that there is no cosmological coupling, confirming previous results in the literature. Furthermore, we show that nonsingular compact objects  couple to the cosmological background, as quantified by their MS mass. We conclude that observational evidence of cosmological coupling of astrophysical BHs would be the smoking gun of their nonsingular nature.}

\begin{document}
\maketitle
\flushbottom

\section{Introduction}{\label{Introduction}}

Recently, there has been renewed interest in an old question of general relativity (GR): are small-scale, local systems, like planets, stars or compact objects/black holes (BHs), affected by the large-scale dynamics of the cosmological background they are embedded in? 

The first known attempt to consistently answer this question dates back to McVittie~\cite{McVittie:1933zz}, who found a solution of Einstein's field equations describing a point-like object embedded in a spatially-flat Friedmann-Lema\^i{}tre-Robertson-Walker (FLRW) spacetime. However, the issue was far from being settled, as the extremely nontrivial physics involved in this embedding entails conceptual and interpretative problems (see, $e.g.$, Refs.~\cite{Sussman:1985,Ferraris:1996ey,Nolan:1998xs,Nolan:1999kk,Faraoni:2012gz}). 

As a result, a considerable body of work has ensued over the years, with, however, contradictory results (for an incomplete list, see, $e.g.$, Refs.~\cite{RevModPhys.17.120,Einstein:1946zz,Pachner:1963zz,dicke1964evolution,Vaidya:1968zza,1971ApJ...168....1N,1981InJPh..55..304G,1982JApA....3...63P,Vaidya:1977zza,DEath:1975jps,Gautreau:1984pny,Cooperstock:1998ny,1990ApJ...360..315H,Buchert:1995fz,Buchert:1999pq,Nayak:2000mr,Baker:2000yh,Bolen:2000dz,Dominguez:2001it,Ellis:2001cq,Gao:2004cr,Sheehan:2004wa,Nesseris:2004uj,Sultana:2005tp,Li:2006zh,Adkins:2006kw,McClure:2006kg,Sereno:2007tt,Faraoni:2007es,Balaguera-Antolinez:2007csw,Mashhoon:2007qm,Carrera:2008pi, Gao:2011tq,Giulini:2013zha,Faraoni:2014nba,Kopeikin:2014qna,Faraoni:2015saa,Mello:2016irl,Faraoni:2018xwo,Macpherson:2018btl,Guariento:2019ock,Neishtadt:2020hcg,Spengler:2021vxy,Agatsuma:2022ewd,Bisnovatyi-Kogan:2023aqf} and references therein). The main conceptual obstacle is caused by the huge separation of scales between local inhomogeneities, whose characteristic scale is their virial radius $\sim GM$, and the large-scale cosmological dynamics, occurring instead at the Hubble radius $H$. Although there seems to be general agreement on the negligible impact of the cosmological expansion on small-scale\footnote{However, this is not necessarily the case with large-scale structures~\cite{1971ApJ...168....1N,Sato:1983ka,Axenides:2000fz,Busha_2003,Balaguera-Antolinez:2007csw,Larena:2008be,Baumann:2010tm,Wiegand:2010uh,Buchert:2015iva,Montanari:2017yki,Adamek:2017mzb,Euclid:2021frk,Schander:2021pgt,Koksbang:2021qqc,Vigneron:2021tpi,DiValentino:2021izs,Abdalla:2022yfr,Zhang:2023neo}.} Newtonian systems~\cite{Iorio:2005vw,Kagramanova:2006ax,Mashhoon:2007qm,Faraoni:2007es}, the issue is still opened for local, relativistic bodies, like BHs. 

Notice that the coupling could be a rather natural feature of highly-compact gravitational systems. Na\"ively, the mass/radius relation of a BH would suggest that, if lengths are affected by the cosmological expansion, so should do masses.

In the latest years, the debate has rekindled due to recent developments based on the theoretical work of Croker and collaborators~\cite{Croker:2019mup, Croker:2020,Croker:2020plg}, 
focusing mainly on supermassive black holes at the center of galaxies as the most promising objects to detect this effect.
Through a perturbative approach and an averaging procedure, they derived the Friedmann's equations from varying the gravitational action. What they show, in this way, is that the pressure in the interior of BHs/compact objects contribute actively to the energy density sourcing the cosmological equations. The conservation of the stress-energy tensor, then, implies the presence of a coupling of these objects with the cosmological expansion, which should manifest as a significant shift in their masses. Their model also allows to predict that the masses of local objects should vary with the scale factor $a$ according to the power law $M(a) \propto a^k$~\cite{Croker:2019mup}. This formula was then tested in Ref.~\cite{Farrah:2023opk} against an observational sample of supermassive BHs at the centre of elliptical galaxies at different redshift. Such objects are quiescent,~$i.e.$,~they are concerned by negligible processes of accretion or mergers, so that the data-set is not sensibly affected by other growth channels other than the supposed coupling mechanism with the expanding background. This set of data showed a preference for $k \sim 3$~\cite{Farrah:2023opk,Cadoni:2023lum}. The conclusion of the authors of Ref.~\cite{Farrah:2023opk} is that BHs may be the source of dark energy. 

However, this claim and the underlying theoretical framework have faced significant criticism. Even if the underlying framework could be flawed from the beginning~\cite{Mistele:2023fds}, most criticism has been directed at the concept of coupling itself. On the one hand, the substantial separation in scales between local and cosmological systems makes such coupling implausible~\cite{Wang:2023aqe,Gaur:2023hmk}. On the other hand, the equation of state of matter inside a BH, which is typically taken to be dust, is unable to mimic dark energy~\cite{Parnovsky:2023wkc,Avelino:2023rac}. 
Moreover, current observational constraints on the slope parameter $k$ capturing the mass-redshift dependence, are highly controversial, as they heavily depend on the astrophysical probes employed in the analysis~\cite{Rodriguez:2023gaa,Andrae:2023wge,Lei:2023mke,Amendola:2023ays}.

These critiques cast again doubts on the feasibility and validity of the proposed coupling between cosmological dynamics and the masses of BHs/compact objects. 
However, in Ref.~\cite{Cadoni:2023lum}, we and collaborators built a solid general relativistic framework that enabled us to describe the coupling of local inhomogeneities with the cosmological background in full generality, as well as to recover the expression of the mass-shift from Refs.~\cite{Croker:2019mup,Farrah:2023opk}.

Apart from the intricate situation on the observational side, the theory behind the cosmological coupling of compact astrophysical objects is far from being well established. One of the main issues is the absence of generic solutions of Einstein's equations describing singular BHs or BH mimickers. This is quite different from the singular BH case, for which we know that spherical, asymptotically-flat solutions are  unique and given by the Schwarzschild one. A simple way to circumvent this problem is to focus on general, model-independent properties, which should characterize the cosmological coupling. Thus, the theoretical question that we will address in the present paper is: in which conditions does the cosmological expansion affect dynamical quantities, such as BH masses? 
We give a precise answer to this question by working on a solid theoretical ground. Previously, the debate was biased by the use of the nonlocal Arnowitt-Deser-Misner (ADM) mass to quantify the energy pertaining to local objects. The starting point of this work, instead, is the identification of the quasi-local Misner-Sharp (MS) mass as the most appropriate quantity to determine the energy of local compact objects, and to investigate their cosmological coupling. 

The MS mass is covariantly defined, and it reduces to the ADM mass at asymptotically-flat infinity. Therefore, it can be identified as an ideal tool for making theoretical predictions to compare with astrophysical measurements.
We then use the MS mass to compute the energy of various cosmologically-embedded solutions: Schwarzschild-de Sitter, McVittie, Sultana-Dyer (SD), and nonsingular BHs sourced by anisotropic fluids.
Additionally, this approach enables us to make a general statement about the existence of the cosmological coupling of compact astrophysical objects. We explicitly show that the cosmological coupling becomes manifest whenever the energy of the central object is quantified by the MS mass. On the other hand, the cosmological decoupling occurs whenever the energy of the central object is equivalent to its ADM mass, as in singular BH models. In particular, this implies that the cosmological coupling is inevitable whenever the energy of the nonembedded object can be quantified everywhere by the MS mass, but not by the ADM one. That is the case of nonsingular BHs and other ultracompact objects.

The paper is structured as follows. 

In \cref{sec:SectionII}, we discuss the role played by the different definitions of mass for BHs/compact objects in cosmology, and provide a brief overview of the key properties of the MS mass.

In \cref{sec:singularembedding}, we revisit the singular Schwarzschild-de Sitter and McVittie solutions, demonstrating that the mass of the central local object does not couple to the cosmological dynamics.

In \cref{sec:nonsingularembedding}, we examine the cosmological embedding of nonsingular BHs/compact objects: we firstly revisit the SD solution, then discuss the most general case of compact objects with anisotropic sources, and finally the isotropic case. We explicitly show that the cosmological coupling is quantified by the MS mass, and we conclude that any observational evidence of cosmological coupling could be a smoking gun for the nonsingular nature of  astrophysical BHs.

We present our conclusions in \cref{sec:conclusions}.

\section{Definitions of mass for compact objects and cosmological coupling}
\label{sec:SectionII}

Answering the question posed in \cref{Introduction}, about the mass growth of compact objects due to cosmological expansion, is complicated by the fact that   there are several definitions of mass/energy in GR \cite{Katz:1996nr,Szabados:2009eka}. Often, quasi-local definitions of energy proposed in the literature are mathematically involved and, hence, difficult to apply to real situations, or even to very idealized and simple analytical solutions of Einstein's equations. On the other hand, the quasi-local MS mass \cite{Misner:1964je} represents a quite natural definition, as it emerges naturally from Einstein's equations and is directly related to astrophysical observations of the internal energy of a spherically-symmetric, virialized system. For isolated objects in asymptotically-flat spacetimes, there is also another relevant definition, which is the ADM mass, a nonlocal quantity defined in terms of a surface integral at spatial infinity.

Static eternal BHs embedded in a cosmological background are usually described by neglecting the cosmological asymptotics and resorting to asymptotic flatness instead, where all observables can be precisely identified and quantified in terms of surface integrals at spatial infinity \cite{Arnowitt:1962hi}. In this case, one can safely use the ADM mass. For spacetimes with different asymptotics, like the FLRW ones, the identification and interpretation of such nonlocal observables become much more involved.

The key issue in this type of problem is not purely kinematic, like, $e.g.$, that concerning the cosmological redshift of distances in cosmology, but fully dynamic. As such, it implies some explicit or implicit assumption about how the small-scale, inhomogeneous dynamics of the compact object is related to the large-scale, homogeneous and isotropic cosmological background dynamics. The usual assumption is that there exists a scale of {\textit{decoupling}}, which is essentially justified by the huge separation of scales between the heaviest known galactic BH  ($\sim 10^{-3}$ pc) and the Hubble radius ($\sim 10^{10}$ pc) \cite{Gaur:2023hmk}.

However, supermassive and stellar-mass BHs have existed for a long time during cosmological evolution. Tiny effects could accumulate over such extremely long time scales, leaving observable  imprints even at small spatial scales. Although the precise form of the cosmological coupling, or its absence for special solutions, cannot be taken as established, the possibility of its presence cannot be \textit{a priori} excluded.

The use of the ADM mass to characterize the energy of cosmologically-embedded BHs is fully justified only if one accepts the assumption of the decoupling of scales. This is the only case where a cosmologically-embedded BH can be safely treated as an eternal, asymptotically-flat object.

The decoupling of scales  assumption is physically justified only for the Schwarzschild-de Sitter solution, where one has a globally-defined, static, radial coordinate to safely define the $r\to 0$ and $r\to \infty$ limits. In other cases, such as,~$e.g.$,~the McVittie solution, the $r\to 0$ and $r\to \infty$ limits use different radial coordinates, related by a time-dependent coordinate transformation.

Finally, another strong limitation of the ADM mass is that it correctly quantifies the energy of the compact objects only in the case of astrophysical bodies in which the stress-energy tensor is zero outside, such as singular BHs. This is not the case for nonsingular BHs~\cite{Cadoni:2022chn}.

In the following, we will briefly review the basic properties of the quasi-local MS mass. 

\subsection{Basic features  of the Misner-Sharp mass}
\label{subsec:MSmass}

Depending on the asymptotics of a given spacetime, there are several ways to quantify the energy of a gravitational system. As already stated, in asymptotically-flat spacetimes (or, more in general, for spacetimes with a  timelike asymptotic boundary, like, $e.g.$, anti de Sitter), the key observables can be unambiguously quantified, through the ADM decomposition, as \textit{nonlocal} quantities defined at the boundary of the spacetime, the so-called ``hair'' of classical, singular Kerr-Newman solutions~\cite{Bardeen:1973gs}. 

For BHs not in vacuum, like nonsingular BHs, this identification is less straightforward, even if the manifold is asymptotically flat (see, $e.g.$, Refs.~\cite{Ma:2014qma,Cadoni:2022chn} and references therein). In these cases, there is a different definition which better encapsulates the \textit{local} properties that the energy of a gravitational system should satisfy: the Hawking-Hayward quasi-local mass~\cite{Hawking:1968qt, Hayward:1993ph}, which, for spherically-symmetric spacetimes, reduces to the MS mass~\cite{Hayward:1994bu}. In a generic asymptotically-flat spacetime, the ADM and MS masses coincide only at spatial infinity, namely $M_\text{ADM} = \lim_{r\to \infty} M_\text{MS}$. They are fully equivalent outside the compact object only if the stress-energy tensor vanishes outside of the object.

On the contrary, the MS mass can be defined \textit{covariantly} also for non-asymptotically flat and non-stationary spacetimes, it is the most natural definition of energy for spherically-symmetric  gravitational systems and is, therefore, routinely used by most researchers working in the field. Moreover, as it encodes the \textit{local} properties of the energy of a given spherically-symmetric, virialized gravitational system, the MS mass is the physical mass measured by astrophysical observations. For these reasons, it represents the most appropriate tool to investigate the possible cosmological coupling of local objects embedded in cosmological backgrounds. 

For a spherically-symmetric spacetime with a metric of the general form\footnote{Throughout the paper, we shall use natural units in which $c=\hbar=k_\text{B}=1$.  Latin indices $a,b=1,2$ denote the time and radial coordinates. We use Greek indices to denote four-dimensional spacetime coordinates. $\dd \Omega$ is the line element of the two-sphere.} 
\begin{equation}
\dd s^2 = h_{ab}(x) \dd x^a \dd x^b + r(x)^2 \dd \Omega^2\, ,
\end{equation}
the MS mass takes the form
\begin{equation}\label{MSmassgeneraldefinition}
M_\text{MS} = \frac{r(x)}{2G}\left[1-h^{ab}(x) \nabla_a r(x) \nabla_b r(x) \right]\, .
\end{equation}

Given that the MS mass is a covariant quantity \cite{Hayward:1994bu}, all the physical results based on its use are coordinate-independent, while its explicit form rests, of course, on the particular gauge chosen. In the following, we consider the systems of coordinates that are mostly adopted when discussing the embedding of spherical objects in cosmological backgrounds. One system is given by Lema\^\i{}tre coordinates $\left(t, \, r, \, \theta, \, \phi \right)$
\begin{equation}
\dd s^2 = -e^{\alpha(t, r)} \dd t^2 + e^{\beta(t, r)} \dd r^2 + R(t, r)^2 \dd \Omega^2\, , \label{metricgeneral1}
\end{equation}
where $\alpha$, $\beta$ and $R$ all depend on the radial and time coordinates. Note also that this metric generalizes the ones written in isotropic coordinates
\begin{equation}
\dd s^2 = -e^{\alpha(t, r)} \dd t^2 + e^{\tilde \beta(t,r)}\left(\dd r^2 + r^2 \dd \Omega^2 \right)\, ,
\end{equation}
that have also been frequently adopted to discuss the cosmological embedding of compact objects. 

However, in order to discuss the MS mass, it is more convenient to use $R(t, \, r)$ as the radial coordinate. Through a straightforward change of coordinates (see Ref.~\cite{Faraoni:2015uma}), one can recast the metric \eqref{metricgeneral1} into the form

\begin{equation}
\dd s^2 = -A(T, R) \dd T^2 + B(T, R) \dd R^2 + R^2 \dd \Omega^2\, , \label{metricgeneral2}
\end{equation}

with the relations
\begin{subequations}
\begin{align}
&A = \left(e^{\alpha}-e^\beta \frac{\dot R^2}{R'^2} \right) F^2\, ; \label{Achangecoordinates}\\
&B = \frac{e^{\alpha + \beta}}{R'^2\left(e^\alpha - e^\beta \frac{\dot R^2}{R'^2} \right)}\, , \label{Bchangecoordinates}
\end{align}
\end{subequations}

where the dot and prime stand for derivation with respect to $t$ and $r$, respectively. $F$ is an integration function entering the time-coordinate transformation, it is required to guarantee that $\dd T$ is an exact differential \cite{Faraoni:2015uma}. Using \cref{metricgeneral2}, the MS mass reads as
\begin{align}
    M_\text{MS} &= \frac{R}{2G}\left(1-g^{\mu\nu}\nabla_\mu R \nabla_\nu R \right) = \frac{R}{2G}\left(1-g^{RR} \right) = \frac{R}{2}\left(1-B^{-1} \right)\, .
\end{align}

Using \cref{Bchangecoordinates}, in the gauge \eqref{metricgeneral1}, it becomes
\begin{equation}\label{MSmassSchwarzschildcoordinates}
M_\text{MS} = \frac{R}{2G}\left(1 + \dot R^2 e^{-\alpha}-R'^2 e^{-\beta} \right)\, .
\end{equation}

In the following sections we will use this formula to compute the mass of cosmologically-embedded compact objects. In order to compute the MS mass for a given spherically-symmetric configuration, we must specify the radius $r_0$ at which $M_\text{MS}$ is evaluated. For horizonless compact objects, like BH mimickers, there is some degree of arbitrariness in the choice of $r_0$  since the density profile always goes to zero only at $r\to \infty$, which prevents from defining a hard surface. We use here the same convention as in Ref. \cite{Cadoni:2023lum}, $i.e.$, we define the surface of the object as the one containing $99\%$  of the MS mass.
Notice that there is no technical complication due to the presence of a time–dependent apparent horizon, which replaces the event horizon for the cosmologically embedded solutions. In fact, as shown in Ref. \cite{Cadoni:2023lum}, the cosmologically embedded solutions we are considering in this paper use spherical spacetime foliations. $r_0$ is only needed at the initial time and, therefore, is given by the radius of the eternal BH event horizon.

\section{Embedding point-like objects and perfect fluid stars in cosmological backgrounds}
\label{sec:singularembedding}

In this Section  we will use the MS mass to reproduce already-known results regarding the nonexistence of a cosmological coupling of standard singular BHs. By doing so, we will clarify the physical reasons behind  the absence of the small-/large- scale coupling in two well-known solutions that were recently reconsidered to advocate against the ubiquity of the cosmological coupling~\cite{Gaur:2023hmk}. The gist of their argument is that there is a complete separation between the scales pertaining to local objects (like BHs) and the dynamics of the cosmological background, such that the mass of the central BH can be approximately identified with its ADM mass. First of all, let us note that the ADM mass cannot be properly defined in non-asymptotically flat spacetimes, as those corresponding to cosmologically embedded objects. Secondly, as previously stressed, the ADM mass is a \textit{nonlocal} quantity, thereby unable to quantify local effects, such as the coupling. Finally, the separation of scales presented in \cite{Gaur:2023hmk} involves rather questionable limits on small and large scales, due to the use of time-dependent radial coordinates. The two limits truly represent separated scales only if we consider the small-scale local dynamics at times for which the scale factor does not change significantly. If this not the case, the notion of small- and large-scale limits would instead depend on time. The only case in which one can truly show that the two limits give neatly separates scales at all times is the Schwarzschild-de Sitter solution, since one can define coordinates in which the spacetime is static.

We shall show that, for point-like objects, the separation of scales and the resulting decoupling emerges naturally when considering the MS mass of the solutions mentioned above.

\subsection{Schwarzschild-de Sitter solution}

The simplest known example of an embedding of a compact object (mass-particle) in a cosmological background is the Schwarzschild-de Sitter metric, which is a vacuum solution of Einstein's equations with a positive cosmological constant. A peculiarity of this metric is that it can be written in the static patch
\begin{equation}\begin{split}
\dd s^2 = &-\left(1-\frac{2Gm}{r} - H^2 r^2 \right) \dd t^2 + \frac{\dd r^2}{1-\frac{2Gm}{r} - H^2 r^2} + r^2 \dd \Omega^2\, .
\end{split}
\end{equation}

As already noted, here we have a clear separation between the small scales, where the solution reduces to the Schwarzschild one with an ADM mass $m$, and the large scales, where we instead have the de Sitter asymptotics. The decoupling can be readily seen using the MS mass given by \cref{MSmassSchwarzschildcoordinates}, instead of a more intricate change of coordinates as done in Ref.~\cite{Gaur:2023hmk}. A straightforward calculation yields indeed
\begin{equation}
M_\text{MS}  
= m + \frac{H^2}{2G} r^3\, .
\end{equation}
The first term is the ADM mass of the Schwarzschild BH, while the second term is simply the mass contribution due to the constant cosmological density over a volume $r^3$. There is no trace of the growth of $m$ due to the expanding cosmological background given by the Hubble parameter $H=\dot a/a$, namely there is no cosmological coupling.

\subsection{McVittie solution}

The McVittie spacetime \cite{McVittie:1933zz} represents a generalization of the Schwarzschild-de Sitter spacetime to a generic FLRW model. It was the first exact solution of GR which allowed for the embedding of spherically-symmetric objects in a generic cosmological background. It is based on some assumptions, the most important ones being a perfect, isotropic and spherically-symmetric fluid as a source, and the absence of fluxes of matter/energy into/away from the central object. Moreover, the metric is required to reduce to the Schwarzschild one, written in isotropic coordinates, when expressed in terms of radial coordinate of the observer, $\hat r= a r$. It thus has the same singularity at the origin. 

In the coordinates used in \cref{metricgeneral1}, it reads as
\begin{equation}\begin{split}\label{macW}
\dd s^2 = &-\frac{\left(1-\frac{Gm(t)}{2 r} \right)^2}{\left(1+\frac{G m(t)}{2r} \right)^2} \dd t^2 + a^2 \left(1+\frac{G m(t)}{2 r} \right)^4 \left(\dd r^2 + r^2 \dd \Omega^2 \right)\, ,
\end{split}\end{equation}

where $a$ is the scale factor and $m(t) = m_0/a(t)$, from Einstein's equations and from the requirement of absence of radial fluxes. 

We identify the areal radius as 
\begin{equation}
R(t, r) \equiv a(t) \, r \left(1+\frac{G m_0}{2 r a(t)} \right)^2\, .
\end{equation}
By writing this solution in the gauge \eqref{metricgeneral2}, \cref{MSmassSchwarzschildcoordinates} yields, 
\begin{equation}\begin{split}
\dd s^2 =& -\left(1-\frac{2Gm_0}{R}-H^2 R^2 \right)F^2 \, \dd T^2 + \frac{\dd R^2}{1-\frac{2Gm_0}{R}-H^2 R^2 } + R^2 \, \dd \Omega^2\, ,
\end{split}\end{equation}
which is very similar to the Schwarzschild-de Sitter solution, the only differences being the factor $F$ in the $g_{TT}$ component, and the fact that $H$ is not restricted to describe a de Sitter cosmological background. This similarity translates also in the behavior of the MS mass, where again we have a complete separation of scales 
\begin{equation}
M_\text{MS} = \frac{R}{2G}\left(\frac{2G m_0}{R} + H^2 R^2 \right) = m_0 + \frac{H^2}{2G} R^3\, .
\end{equation}

We again identify two independent terms: a contribution of the ADM mass of the BH, and a purely cosmological term. No coupling is present, as expected, since the density sourcing the McVittie solution is purely cosmological,~$i.e.$, $\rho = 3H^2/8\pi G$, while the contribution of the ADM mass of the central object can only be accounted for by inserting by hand the usual Dirac delta distribution. 

\section{Coupling of compact objects: local anisotropic and isotropic sources}
\label{sec:nonsingularembedding}

In \cref{sec:singularembedding} we showed that, for solutions of Einstein's equations describing the cosmological embedding of the Schwarzschild black hole, there is no cosmological coupling. 

On the contrary, when the impact of small-scale anisotropies is taken into account, it is possible to have nontrivial solutions describing compact objects/BHs \cite{cosenza1981some,Herrera:1997plx,Raposo:2018rjn} circumventing the Penrose theorem~\cite{Penrose:1964wq}, and enabling the construction of nonsingular solutions \cite{bardeen1968proceedings,Dymnikova:1992ux,Bonanno:2000ep,Modesto:2004xx,Hayward:2005gi,Ansoldi:2008jw,Modesto:2008im,Nicolini:2008aj,Spallucci:2011rn,Frolov:2016pav,Fan:2016hvf,Simpson:2019mud,Lobo:2020ffi,Cadoni:2022chn,Akil:2022coa}. 
In this Section, we will consider several scenarios of cosmological embedding of compact objects: the SD solution, solutions sourced by anisotropic fluids, solutions sourced by isotropic fluids, and charged singular solutions.\\ 

\subsection{The Sultana-Dyer solution}{\label{subsec:SD}}
The SD solution~\cite{Sultana:2005tp} is another exact solution describing a BH embedded in a spatially flat FLRW. It was found by conformally transforming the Schwarzschild metric with the goal of changing the Schwarzschild global timelike Killing vector into a conformal Killing one. The conformal transformation also allows the spacetime to be nonsingular at $r=0$. 

Despite it being problematic due to the fluid becoming tachyonic at late times near the horizon~\cite{Sultana:2005tp,Faraoni:2007es}, it is still interesting for our purposes. 

The metric is essentially the McVittie one \eqref{metricgeneral1}, with an important difference: the mass $m$ appearing in the metric is now a constant 
\begin{equation}\label{ghj}
\dd s^2 = -\frac{\left(1-\frac{G m_0}{2r} \right)^2}{\left(1+\frac{G m_0}{2r} \right)^2} \dd t^2 + a^2 \left(1+\frac{G m_0}{2r} \right)^4 \left(\dd r^2 + r^2 \dd \Omega^2 \right)\, .
\end{equation}

The fact that the mass does not depend on $a$ naturally introduces fluxes, making the source anisotropic. In fact, the source of the SD solution is a combination of two noninteracting perfect fluids, one in the form of an ordinary massive dust and the other of a null dust \cite{Sultana:2005tp,Faraoni:2007es}. It is well-known that such a combination can be recast as a single anisotropic fluid \cite{Bayin:1985cd}. As we shall see, this is the origin of the coupling with the cosmological background.  

We now compute the MS mass \eqref{MSmassSchwarzschildcoordinates} of the SD solution. We first identify 
\begin{equation}
R \equiv a r \left(1+\frac{G m_0}{2r} \right)^2\, ,
\end{equation}
with which, using \cref{Bchangecoordinates} and reading $e^\alpha$ and $e^\beta$ from \cref{ghj}, we get
\begin{equation}
B 
= \frac{1-\frac{2Gam_0}{R}}{\left(1-\frac{2Ga m_0}{R} \right)^2-H^2 R^2}\, .
\end{equation}
Therefore, the MS mass \eqref{MSmassSchwarzschildcoordinates} reads as
\begin{equation}\label{sultana}
M_\text{MS} 
= a \, m_0 + \frac{H^2 R^3}{2G \left(1-\frac{2Ga m_0}{R} \right)}\, .
\end{equation}
 
The first term represents the coupling of the mass of the solution with the cosmological background, and it is consistent with the linearly-scaling universal coupling term derived for the first time in Ref.~\cite{Cadoni:2023lum} for generic anisotropic fluids.
The second term cannot be interpreted as a pure cosmological contribution, due to the presence, in the denominator, of a term depending on $a \, m_0$. The latter encodes the interaction between the small and large scales, as a physical consequence of the accretion flow of cosmic fluid onto the central object, due to the presence of nonzero fluxes in the source. Note that similar results hold for the class of exact models analyzed in Ref.~\cite{Faraoni:2007es}, devised to correct the problems of the SD metric, as well as for the solutions considered in Ref.~\cite{Culetu:2012xeh}.

\subsection{Compact objects sourced by anisotropic fluids}
\label{subsec:anf}
In Ref. \cite{Cadoni:2023lum} it was shown that the metric parametrization ($\eta$ is the conformal time)
\begin{equation}\label{metricourwork}
\dd s^2 = a^2(\eta)\left[-e^{\alpha(\eta, r)} \dd \eta^2 + e^{\beta(\eta, r)} \dd r^2 + r^2 \dd\Omega^2\right]\, ,
\end{equation}
representing the cosmological embedding of a generic compact object sourced by an anisotropic fluid, allows to describe the coupling of GR BHs/horizonless configurations to the cosmological background. The stress-energy tensor pertaining to the source has the form $T^{\mu}_\nu = \text{diag}\left(-\rho, \, p_{\varpar}, \, p_{\perp}, \, p_{\perp} \right)$.

Einstein's equations and stress-energy tensor conservation give:
\begin{subequations}\label{isofluid}
\begin{align}
&e^{-\beta(r,\eta)}= g(r) a^{r\alpha'}\, ; \label{betafunction}\\
&\frac{\dot{a}^2}{a^2}\left(3-r\alpha' \right)e^{-\alpha} + \frac{1-e^{-\beta}+r\beta'e^{-\beta}}{r^2} = 8\pi Ga^2 \rho\, ; \label{alpha00}\\
&\frac{e^{-\beta}+re^{-\beta} 	\alpha'-1}{r^2}+e^{-\alpha}\left(-2\frac{\ddot{a}}{a}+\frac{\dot{a}^2}{a^2} \right)=8\pi G a^2 p_\parallel\, ; \label{alpharr}\\
& \dot{\rho} + \frac{\dot{a}}{a}\left(3\rho+3p_{\varpar}+rp'_{\varpar} \right)=0\, , \label{alphaconserv}
\end{align}
\end{subequations}
where a dot now means derivation with respect to $\eta$. The remaining equation, stemming from the conservation of the stress-energy tensor, is  used to compute $p_\perp$.

The field equations allow for a regime in which $\dot \alpha = 0$, which is the only one in which compact objects can be consistently embedded in a cosmological background (see Refs.~\cite{Cadoni:2020jxe,Cadoni:2021zsl,Cadoni:2023lum}). It describes the \textit{absence} of fluxes, unlike the case analyzed in \cref{subsec:SD}.

This set-up is suitable to describe both the large- (cosmological) and the small- (inhomogeneity) scale dynamics, and it allows for a non-zero interaction term between these two scales\footnote{It is worth noting that the cosmological embedding realized using the metric \eqref{metricourwork} is more general than the one used by McVittie in \cref{macW}. In our case the metric is required to reduce to the static metric of the local compact object at any fixed instant of time. Conversely, in the McVittie parametrization, such reduction must happen by passing to the observer radial coordinate $\hat r= a r$.}.

The cosmological mass coupling is immediately manifest when computing the density through Einstein's equations and integrating it over a reference cosmological volume (see Ref.~\cite{Cadoni:2023lum} for further details)
\begin{equation}\begin{split}
M(\eta) &= 4\pi a^3(\eta) \int_0^L \dd r \, r^2 \, \rho(r, \eta)\\
&= \frac{4\pi }{3}\rho_1 a^3 L^3 \, e^{-\alpha(L)} + M(a_i) \frac{a}{a_i} \left[1-e^{-\beta_0(L)}a^{k_L} \right]\, ,
\label{MSmasscosmologygeneral}
\end{split}\end{equation}
where $L$ is the scale of a particular compact objects, while 
\begin{equation}
    k_L \equiv k(L) = r \alpha'(r) \biggr|_{r = L} \, .    
\end{equation}
Here we defined $M(a_i) \equiv a_i L/2 G$, which is the mass of the object computed at the coupling epoch. This expression defines the proper Schwarzschild radius $a_i L = 2G M(a_i)$ at this reference time \footnote{The definition of $M(a_i)$ as a linear function of $L$ is expected to hold only for black holes or ultra-compact, strongly-coupled, objects, for which a weak-field limit description is not valid.}. 

The first term in \cref{MSmasscosmologygeneral} corresponds to a purely cosmological contribution, which depends on the cosmological background energy density $\rho_1$. It is therefore expected to be relevant only beyond the transition scale to homogeneity and isotropy. It does not play a role at the typical scales $L$ of the compact object. As here we are not interested in whether and how  the small-scale dynamics affects the large-scale cosmological one, another long-standing problem commonly known as "cosmological backreaction" \cite{Larena:2008be,Baumann:2010tm,Wiegand:2010uh,Buchert:2015iva,Montanari:2017yki}, we can safely neglect this term in the following discussion.

The second term in \cref{MSmasscosmologygeneral} represents a ``universal cosmological Schwarzschild mass". The cosmological coupling emerges as a linear dependence between the mass of the object and the scale factor $a$. Note that this is the same universal coupling term found for the SD solution (see \cref{sultana}). It has here a geometric origin in terms of the local curvature generated by the compact object~\cite{Cadoni:2023lum}. Finally, the last term encodes model-dependent corrections to the universal term. 

Note that, for standard Schwarzschild BHs, the sum of the second and third terms in \cref{MSmasscosmologygeneral} is identically zero, and we are left with the purely cosmological contribution. 

Eq.~\eqref{MSmasscosmologygeneral} can also be derived from the general definition of the MS mass. Using $R = a r$, from \cref{MSmassSchwarzschildcoordinates} one gets

\begin{equation}\begin{split}
M_\text{MS} &= \frac{a r}{2G}\left[1 + \frac{\dot a^2}{a^2} r^2 e^{-\alpha} -\frac{a^2}{a^2}e^{-\beta} \right]\\
&= \frac{4\pi}{3}\rho_1 a^3 r^3 e^{-\alpha} + \frac{ar}{2G}\left[1-e^{-\beta_0(r)} a^{k(r)} \right]\, ,
\label{MSourmodel}
\end{split}\end{equation}

where, in the last step, we made use of the Friedmann equation $3 \dot a^2/a^2 = 8\pi G a^2 \rho_1$, with $\rho_1 = \rho_1 (\eta)$ being the density of the cosmological fluid sourcing the background. To obtain the last element we also used \cref{betafunction}. 
\cref{MSourmodel} is equivalent to \cref{MSmasscosmologygeneral} when evaluated at the radius of the compact object $r = L$. 

Notice that the cosmological coupling is present independently of the underlying cosmological background (for instance, it is present also in a de Sitter background). That is because $\rho_1$ can be freely specified, and it determines the scale factor \textit{regardless} of the dynamics of the small-scale inhomogeneities\footnote{This is strictly true only if one neglects the backreaction (see the discussion above)}. If we impose the solution at constant time to be Schwarzschild-de Sitter, we have no coupling,~$i.e.$,~we have a complete separation between scales. Let us finally recall that, differently from the SD and other solutions, here the coupling is not due to some accretion onto compact objects, since we imposed the absence of radial fluxes.

\subsection{Compact objects sourced by isotropic fluids}
The case of nonsingular compact objects sourced by isotropic fluids can be  considered as a particular case of the previously discussed anisotropic fluid, where $p_\perp=p_\varpar=p$. The only difference is that, now, the conservation of the stress-energy tensor gives the additional equation $p^{\prime} + \alpha^{\prime}\left( \rho + p \right)/2 = 0$. This makes the system \eqref{isofluid} more constrained, so that, as shown by McVittie~\cite{McVittie:1933zz}, it does not allow for smooth solutions describing the cosmological embedding of compact objects. Only the standard FLRW cosmological solutions are allowed. On the contrary, in the derivation of the main results of \cref{subsec:anf}, $i.e.$, \cref{MSmasscosmologygeneral,MSourmodel}, we did not exploit the conservation equations for the stress-energy tensor. Therefore they still hold true also in the case of isotropic compact objects, provided that the cosmologically embedded solutions exist. This could be, for instance, the case of a non-smooth cosmological embedding of a local compact object. Thus, we expect also nonsingular compact objects sourced by isotropic fluids to couple to the cosmological evolution in the same way as their anisotropic counterparts.

\subsection{Charged singular solution embedded in a FLRW background}

The cosmological embedding of singular solutions can also be extended to the charged Reissner-Nordstr\"om (RN) solution of GR. There has been some work devoted to the charged generalization of the McVittie spacetime, the so-called Shah-Vaidya solution (see Refs.~\cite{Vaidya:1968zza,Gao:2004cr,McClure:2006kg,Faraoni:2014nba,Guariento:2019ock} and references therein). Even in this case, the stress-energy tensor is anisotropic, but any flux onto/away from the central object is absent.
One might attempt to na\"i{}vely apply the general results of this section  to the cosmological embedding of the RN BH,  given that the stress-energy tensor is nonzero outside of the horizon, 
implying that the MS and ADM mass are not identical in the BH exterior.
However, there is a problem in the definition of the quasi-local mass, due to the divergence associated to the electric field at $r=0$.

In the presence of an electric field, the gravitational potential scales as $1/r^2$, due to the density of the electric field scaling as $r^{-4}$. As a consequence, there is an extra factor -- scaling as $1/r$ in the quasi-local energy -- which is ill-defined at $r = 0$. This feature is interpreted as a repulsive effect due to the intensity of the electric field in the vicinity of the singularity (for further details, see, $e.g.$, Section VIII in Ref.~\cite{Hayward:1994bu}, or Section III in Ref.~\cite{VChellathurai_1990}).  

Alternatively, one can quantify the energy of a charged solution with the ADM mass, which is anyhow undefined in non-asymptotically flat spacetimes, and moreover does not capture the coupling to the cosmological background, as discussed in detail throughout this paper. We thus conclude that a clear description of the coupling between a charged singular spacetime embedded in a cosmological background is quite problematic. Moreover, from a purely phenomenological point of view, the presence of an electromagnetic charge is astrophysically irrelevant.

\section{Conclusions}
\label{sec:conclusions}

In this paper we have shown that the longstanding debate on the theoretical status of the coupling of BHs/compact astrophysical objects to the cosmological background can be finally put on solid ground, if the physically observable mass of the compact object is identified as the MS mass. By doing so, we have explicitly demonstrated that singular BHs cannot couple to the large-scale cosmological dynamics.

We have also shown that the cosmological coupling is not only allowed, but quite natural, for generic compact objects sourced both by isotropic fluids and local anisotropies, like, $e.g.$, nonsingular BHs. The energy of these systems has a quasi-local nature, so that it can be correctly quantified by the MS mass (instead of the nonlocal ADM mass). 
In this case, we have found that the mass of the object is intrinsically linked to the scale factor $a$ by a universal linearly-scaling leading term, with a geometric origin in terms of the local curvature of spacetime. 
Although our derivation provides a general, model-independent statement on the existence of the cosmological coupling of
compact astrophysical objects, its range of application is until now quite limited. This is because, up to now, there is no generic solution for BHs or BH mimickers to be embedded in a FLRW universe. Our explicit examples of regular BHs and BH mimickers are, therefore, limited to the few cases already present in the literature. On the other hand, our considerations could be straightforwardly applied to any future new solution of this kind.

From a purely theoretical point of view, there are two main issues that still need further investigation: $(i)$ fully understand the cosmological coupling of singular objects sourced by local anisotropies, but characterized by an ill-defined MS mass (which prevents a direct application of the results of \cref{sec:nonsingularembedding}). This is  the case of the  RN BH briefly discussed in this paper; $(ii)$ generalize the framework in order to encompass rotating compact objects (see, $e.g.$, Refs.~\cite{1981InJPh..55..304G,1982JApA....3...63P,Vaidya:1977zza} for earlier works on the embedding of the Kerr metric in a FLRW cosmology). Note, however that including rotation is expected to have a non-neglibile impact only on the subleading non-universal term in \cref{MSmasscosmologygeneral}, without affecting at all our conclusions about the leading universal linear coupling term and the decoupling of singular BHs.

As described in \cref{Introduction},  the situation still remains rather   complicated from the point of view of  observations, mainly due to the lack of clear observational results.
New sets of data are needed in order to validate the theoretical predictions for the exponent $k$. Our theoretical analysis shows quite clearly that eternal BHs, $i.e.$, objects with event horizons, are characterized either by $k=0$ if they are singular objects, or by $k=1$ if they are regular. It seems that GR can not allow for other possibilities.
Observational evidence of a nonzero cosmological coupling would thus be the smoking gun of the nonsingular nature of the actual astrophysical black holes. Conversely, a clear detection of $k = 0$ would imply that nonsingular GR BHs are hardly compatible with observations.

\section*{Acknowledgements}
We thank Luca Amendola, Valerio Faraoni, Francesca Lepori and Davi C. Rodrigues for very interesting and fruitful discussions.

\bibliography{refs}
\end{document}